\documentclass[preprint,proceedings]{rmaa}
\usepackage{paralist}

\usepackage{psfrag,color}

\SetYear{2003}
\SetConfTitle{IAUC 194: Compact Binaries in the Galaxy and Beyond}

\title{The crust cooling curve of the neutron star in MXB 1659--29}

\author{
  Rudy Wijnands\altaffilmark{1}} 
\altaffiltext{1}{School of Physics \& Astronomy, U. of St Andrews, UK.}
\shortauthor{R. Wijnands}
\shorttitle{Neutron-star crust cooling curve}

\fulladdresses{
\item Rudy Wijnands:  School of Physics \& Astronomy, University of 
St Andrews, North Haugh, St Andrews, Fife, UK
(\email{radw@st-andrews.ac.uk}). Present address: Astronomical 
Institute 'Anton Pannekoek', University of Amsterdam, Kruislaan 403,
1098 SJ, Amsterdam, The Netherlands (\email{rudy@science.uva.nl}).

}

\listofauthors{R. Wijnands}
\indexauthor{Wijnands, R.}

\abstract{
We have monitored the quasi-persistent neutron-star X-ray transient
MXB 1659--29 in quiescence using {\itshape Chandra}. The purpose of
our observations was to monitor the quiescent behavior of the source
after its last prolonged outburst episode and to study the cooling
curve of the neutron star in this system. We discuss the results
obtained and how they constrain the properties of the neutron star in
MXB 1659--29.}

\resumen{
Hemos observado repetidamente a MXB 1659--29, un sistema transitorio
de rayos X que presenta extensos per\i\'odos de transferencia de masa
(fuente ``cuasi-persistente'' de rayos X). La meta de nuestras
observaciones es documentar el comportamiento de este objeto despue\'s
de su u\'ltimo per\i\'odo extenso de transferencia de masa y de esta
manera estudiar la curva de enfriamiento de la estrella de neutrones
en este sistema. En este art\i\'culo presentamos nuestros resultados
asi como una discusio\'n de las implicaciones que e\'stos tienen para
las propiedades de la estrella de neutrones en MXB 1659--29.

}

\addkeyword{stars: neutron stars: individual (MXB 1659--29)}
\addkeyword{X-rays: stars}

\begin{document}
\maketitle

\section{Introduction}
\label{sec:intro}

Neutron-star X-ray transients spend most of their time in a quiescent
state during which hardly any or no accretion takes place. However,
occasionally they can become very bright in X-rays due to a huge
increase in the accretion rate onto their neutron stars. Most
neutron-star transients are only active for several weeks to a few
months at most, but several systems have been found to be active for
years and even decades. Those systems have been called
'quasi-persistent' neutron-star X-ray transients.

During an outburst of the 'ordinary' transients, the neutron-star
crust is only marginally heated. However, in the quasi-persistent
transients the neutron star is significantly affected by the accreted
material (e.g., Wijnands et al.~2001; Rutledge et al.~2002). In
particular, the neutron-star crust is heated to high temperatures and
will become considerably out of thermal equilibrium with the core
(Rutledge et al.~2002). After the end of the outbursts, the crust will
cool until it is again in thermal equilibrium with the core. The
cooling time depends on the microphysics of the crust and the core,
and the accretion history of the source (Rutledge et al.~2002).

Wijnands et al.~(2001) observed the quasi-persistent transient KS
1731--260 with {\itshape Chandra} within a few months after its last
outburst (which lasted for $\sim$12.5 years). Wijnands et al.~(2002)
reported on an {\it XMM-Newton} observation of the source taken
$\sim$6 months after this {\it Chandra} observation and found that
within half a year the source had decreased in 0.5--10 keV flux by a
factor of $\sim$3. Rutledge et al.~(2002) have calculated crust
cooling curves for the neutron star in KS 1731--260 and based on those
curves Wijnands et al.~(2002) concluded that the neutron star in this
system must have a large thermal conductivity in its crust and
enhanced core cooling processes. In September 2001, a second
quasi-persistent transient (MXB 1659--29) turned off. We initiated a
monitoring campaign using {\itshape Chandra} to study the crust
cooling curve of the neutron star in this system. Here we briefly
describe the results of this campaign (for a detailed discussion see
Wijnands et al.~2003, 2004).

\section{Observations and results}

\begin{figure}[!t]\centering
  \includegraphics[angle=-90,width=0.9\columnwidth]{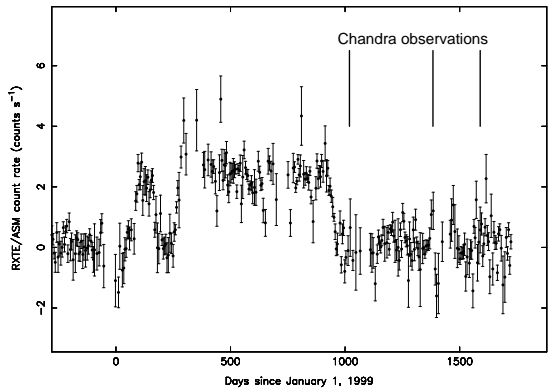}
\caption{The ASM count-rate curve of MXB 1659--29 showing
the $\sim$2.5 year outburst. The times of the {\it Chandra}
observations are indicated by the vertical lines}
\label{fig:asm}
\end{figure}

The quasi-persistent neutron-star X-ray transient MXB 1659--29 was
active in the mid 1970's for $\sim$2.5 years after which it remained
quiescent until April 1999 (in 't Zand et al.~1999; see Wijnands et
al.~2003 for an overview of the accretion history of MXB
1659--29). Again the source remained active for $\sim$2.5 years until
September 2001. Figure~\ref{fig:asm} shows the X-ray count-rate curve
as obtained with the all-sky monitor (ASM) aboard the {\it Rossi X-ray
Timing Explorer} ({\it RXTE}) showing this recent outburst (see also
Wijnands et al. 2003).  As explained above (\S \ref{sec:intro}) this
2.5 year outburst episode should have heated the neutron-star crust
out of thermal equilibrium with the core and, once back in quiescence,
the crust should slowly cool until it comes to thermal equilibrium
again with the core. Therefore, we initiated a monitoring campaign
with {\itshape Chandra} to study the cooling curve of this source. The
details of those {\itshape Chandra} observations are given in Wijnands
et al.~(2003, 2004). The observations were taken $\sim$1, $\sim$12,
and $\sim$19 months after the end of the prolonged outburst (see
Fig.~\ref{fig:asm}).

\begin{figure}[!t]
  \includegraphics[width=\columnwidth]{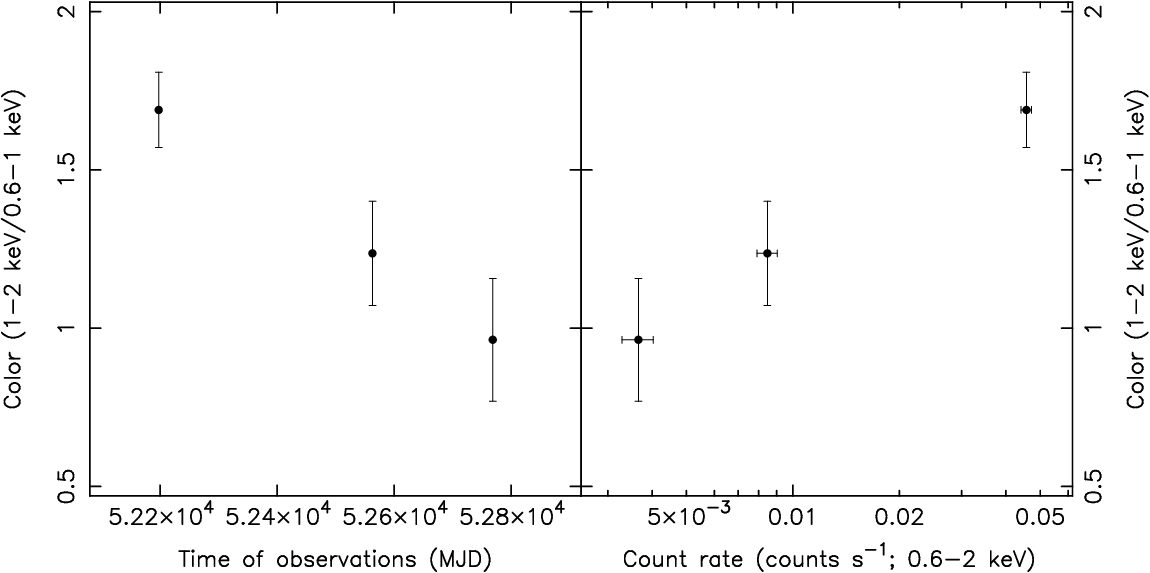}
\caption{The color vs. time (left) and  color vs. the count rate  (right)
during those observations.}
\label{fig:colors}
\end{figure}

For each observation we obtained the background-corrected count rates
and calculated a X-ray color (see Wijnands et al.~2004 for details
about the extraction method and regions). We used the 0.6--2 keV count
rate and the ratio of the 1--2 keV count rate to the 0.6--1 keV count
rate as X-ray color. Those particular energy ranges were chosen
because the source was detected during each observations in the energy
range 0.6--2 keV (i.e., during the last observation hardly any source
photons were found outside this range). The 0.6--2 keV count rate
decreased significantly with time (by a factor of $\sim$10; Wijnands
et al.~2004).  In Figure \ref{fig:colors} we show the behavior of the
color with respect to time and with respect to the 0.6--2 keV count
rate. As the source decreased in luminosity (i.e., as the quiescence
time increased), its spectrum became softer. This softening cannot be
explained by the decrease in time of the ACIS quantum efficiency (the
colors are not corrected for this) since this degradation occurs
mostly at soft energies. If the colors were corrected for this
degradation they would become even lower in time.

\begin{figure*}[!t]\centering
  \includegraphics[angle=-90,width=0.9\textwidth]{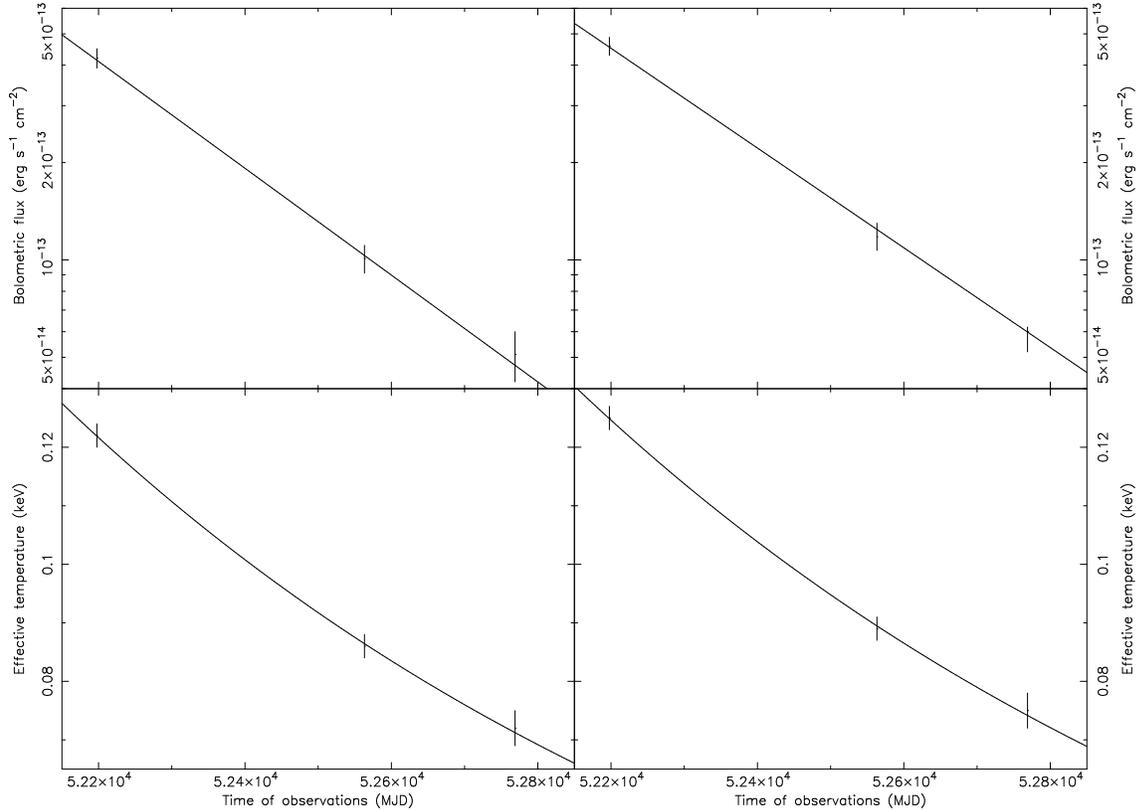}%
\hspace*{\columnsep}%
\caption{
The bolometric flux (top panels) and effective temperature
curves (bottom panel; for an observer at infinity). The left panels
are obtained by using the NSA model of Zavlin et al.~(1996) and the
right panels by using that of G\"ansicke et al.~(2002) (assuming a
hydrogen atmosphere and for a non-magnetized neutron star).}
\label{fig:curves}
\end{figure*}

If the observed X-rays are due to emission from the neutron-star
surface, then the softening of the spectra with time would suggest
that the temperature of the neutron-star crust decreased considerably
in time. To test this hypothesis, we extracted the X-ray spectrum for
each observation (see Wijnands et al.~[2004] for details). The
obtained spectra (corrected for background and ACIS efficiency
degradation) are shown in Wijnands et al.~(2004). The spectra were
fitted with a neutron-star atmosphere (NSA) model for non-magnetized
neutron stars (i.e., that of Zavlin et al.~[1996] and G\"ansicke et
al.~[2002]; assuming a canonical neutron star with a mass of 1.4 solar
masses and a radius of 10 km; the distance toward the source was
assumed to be 10 kpc). The decrease in count rate and the softening of
the spectra with time can be explained by a decrease in bolometric
flux due to a decrease in effective temperature. In Figure
\ref{fig:curves} we show the bolometric flux (top panels) and
effective temperature (bottom panels) of the neutron-star crust as a
function of time. The left panels were obtained by using the Zavlin et
al.~(1996) NSA model (see also Wijnands et al.~2004) and the right
panels by using the G\"ansicke et al.~(2002) model (which could not be
displayed by Wijnands et al.~2004 because of space limitations).

When using the G\"ansicke et al.~(2002) NSA model we get slightly
higher temperatures (and consequently also higher bolometric
luminosities) than when using the Zavlin et al.~(1996) NSA model. This
is a known discrepancy between the two models and the reason for this
is not understood (G\"ansicke et al.~2002). However, no significant
differences are found in the way the bolometric flux and the effective
temperature decrease with time: for both curves the decrease can be
described as an exponential decay function with $e$-folding times of
262$\pm$33 and 282$\pm$19 days for the bolometric flux and
1060$\pm$126 and 1096$\pm$129 days for the effective temperature (for
the Zavlin et al.~[1996] and the G\"ansicke et al.~[2002] models
respectively).

\section{Conclusion}

Our results show that in its quiescent state MXB 1659--298
significantly decreased in brightness within 1.5 years after the end
of its last prolonged outburst. Along with this brightness
decrease, the source became progressively softer, independently of which
model is fit to the spectral data. When the X-ray spectra obtained are
fitted with a NSA model, this brightness decrease and softening of the
spectrum can be explained as a decrease in effective temperature of
the neutron-star surface. Wijnands et al.~(2004) suggested that we see
the neutron-star crust cool in time and that the fast cooling time
might suggest that the neutron-star crust has a large thermal
conductivity and that enhanced core cooling processes occur in the
core. Detailed cooling curves specifically calculated for MXB 1659--29
(similar to those calculated for KS 1731--260 by Rutledge et al.~2002)
are needed to determine whether these conclusions will hold and what
the exact impact of our observations are for our understanding of the
neutron-star properties in MXB 1659--29.

\end{document}